\documentclass[twocolumn,showpacs,preprintnumbers,amsmath,amssymb,prl,superscriptaddress]{revtex4}
\usepackage[T1]{fontenc}
\usepackage[latin1]{inputenc}
\usepackage{graphicx}

\usepackage{color}

\newcommand{\beq}{\begin{equation}}
\newcommand{\eeq}{\end{equation}}

\newcommand{\beqa}{\begin{eqnarray}}
\newcommand{\eeqa}{\end{eqnarray}}
\newcommand{\intd}{\textrm d}
\newcommand{\intR}{\int_{-\infty}^\infty}

\newcommand{\etal}{\mbox{\textit{et al.}}}

\newcommand{\IV}{\mbox{$I$--$V$}}

\newcommand{\Eqref}[1]{Eq.~(\ref{#1})}

\makeatother

\begin{document}

\title{Efficient calculation of inelastic vibration signals in electron transport:\\ Beyond the wide-band approximation}
\author{Jing-Tao L\"u}
\affiliation{School of Physics, Huazhong University of Science and Technology, Wuhan, China}
\affiliation{Dept. of Micro- and Nanotechnology, Technical University of Denmark, {\O}rsteds Plads, Bldg.~345E, DK-2800 Kongens
Lyngby, Denmark}
\author{Rasmus B. Christensen}
\affiliation{Dept. of Micro- and Nanotechnology, Technical University of Denmark, {\O}rsteds Plads, Bldg.~345E, DK-2800 Kongens
Lyngby, Denmark}
\author{Giuseppe~Foti}
\affiliation{Centro de F\'isica de Materiales, Centro Mixto CSIC-UPV, Donostia-San Sebasti\'an, Spain}
\affiliation{Donostia International Physics Center (DIPC) -- UPV/EHU, Donostia-San Sebasti\'an, Spain}
\author{Thomas~Frederiksen}
\affiliation{Donostia International Physics Center (DIPC) -- UPV/EHU, Donostia-San Sebasti\'an, Spain}
\affiliation{IKERBASQUE, Basque Foundation for Science, Bilbao, Spain}
\author{Tue Gunst}
\affiliation{Dept. of Micro- and Nanotechnology, Technical
University of Denmark, {\O}rsteds Plads, Bldg.~345E, DK-2800 Kongens
Lyngby, Denmark}
\author{Mads~Brandbyge}
\affiliation{Dept. of Micro- and Nanotechnology, Technical
University of Denmark, {\O}rsteds Plads, Bldg.~345E, DK-2800 Kongens
Lyngby, Denmark}
\email{mads.brandbyge@nanotech.dtu.dk}
\date{\today}

\begin{abstract}
We extend the simple and efficient lowest order expansion (LOE) for inelastic electron tunneling spectroscopy (IETS) to 
include variations in the electronic structure on the scale of the vibration energies. This enables first-principles 
calculations of IETS lineshapes for molecular junctions close to resonances and band edges. We demonstrate how this is relevant for 
the interpretation of experimental IETS using both a simple model and atomistic first-principles simulations.
\end{abstract}

\maketitle

The inelastic scattering of electronic current on atomic vibrations is a
powerful tool for investigations of conductive atomic-scale junctions.
Inelastic electron tunneling spectroscopy (IETS) has been used to probe molecules
on surfaces with scanning tunneling
microscopy (STM) \cite{StReHo.98.Single-moleculevibrationalspectroscopy}, and
for junctions more symmetrically bonded between the
electrodes \cite{AgUnRu.02.Onsetofenergy,SmNoUn.02.Measurementofconductance,KuLaPa.04.Vibroniccontributionsto,YuKeCi.04.Inelasticelectrontunneling,Song2009,Okabayashi2013}.
Typical IETS signals show up as dips or peaks in the second derivative of the
current-voltage (\IV) curve \cite{GaRaNi07}.  In many cases the bonding geometry is unknown in
the experiments. Therefore, first-principles transport calculations at the level of
density functional theory (DFT) in combination with nonequilibrium Green's
functions (NEGF) \cite{SeRoGu.05.Abinitioanalysis,PaFrBr.05.Modelinginelasticphonona,JiKuLu.05.First-principlessimulationsof,SoGaPe.06.Understandinginelasticelectron-tunneling,Frederiksen2007,Rossen13}
can provide valuable insights into the atomistic structure and IETS.
For systems where the electron-vibration (e-vib) coupling is
sufficiently weak and the density of states (DOS) varies slowly with energy (compared to
typical vibration energies) one can greatly simplify calculations
with the lowest order expansion (LOE) in terms of the e-vib coupling
together with the wide-band
approximation (LOE-WBA) \cite{PaFrBr.05.Modelinginelasticphonona,ViCuPa.05.Electron-vibrationinteractionin}.
The LOE-WBA yields simple expressions for the inelastic signal in terms of
quantities readily available in DFT-NEGF calculations. Importantly, the LOE-WBA
can be applied to systems of considerable size.

However, the use of the WBA can not account for IETS signals close to electronic resonances or band edges, 
which often contains crucial information \cite{PEBA.87.INELASTICELECTRON-TUNNELINGFROM,Egger08}. For example, 
a change in IETS signal {\em from peak to peak-dip} shape was recently reported by Song \etal~\cite{Song2009} 
for single-molecule benzene-dithiol (BDT) junctions, where an external gate enabled tuning of the transport 
from off-resonance to close-to-resonance. Also, high-frequency vibrations involving hydrogen appear problematic 
since the LOE-WBA is reported to underestimate the IETS intensity \cite{Okabayashi2010}. 

Here we show how the energy dependence can be included in the LOE description without changing significantly the transparency of the formulas or the computational cost.
We describe how the generalized LOE differs from the original LOE-WBA, and demonstrate that it captures the IETS lineshape close to a resonance.
We apply it to DFT-NEGF calculations on the resonant BDT system and to off-resonant alkane-dithiol junctions, and show how the improved LOE is necessary to explain the experimental data.

{\it Method.} We adopt the usual two-probe setup with quantities defined in a local basis set in the central region ($C$) coupled to left/right electrodes ($\alpha=L,R$).
We consider only interactions with vibrations (indexed by $\lambda$ with energies $\hbar\omega_\lambda$ and e-vib coupling matrices ${\bf M}_\lambda$) inside $C$. 
To lowest order in the e-vib self-energies ${\bf \Sigma}_{\lambda}$ (2nd order in ${\bf M}_\lambda$) the
current can be expressed as a sum of two terms, $I(V)=I_e+I_i$, using unperturbed Green's functions $\mathbf G^a={\mathbf G^r}^\dagger$ defined in region $C$
\cite{PaFrBr.05.Modelinginelasticphonona,ViCuPa.05.Electron-vibrationinteractionin},
\begin{eqnarray}
\label{eq:MeirWingreenCurrent2}
I_e &=& \frac{\mathrm{G}_0}e \intR \!\!{\!\intd \varepsilon} \left\{ f_L(\varepsilon)-f_R(\varepsilon) \right\} \left\{
\textrm{Tr}[\mathbf G^r \mathbf\Gamma_L \mathbf G^a\mathbf\Gamma_R](\varepsilon) \right. \nonumber\\
&& \qquad\qquad\left. + 2 \textrm{Re\,Tr}[\mathbf G^r {\bf \Sigma}_{\lambda}^{r} \mathbf G^r\mathbf\Gamma_L \mathbf G^a\mathbf\Gamma_R](\varepsilon) \right\},\label{eq.Iel}\\
I_i&=&\frac{\mathrm{G}_0}e \intR \!\!{\!\intd \varepsilon}
\textrm{Tr}[\mathbf\Sigma^{<}_{\lambda}\mathbf G^{r} \mathbf\Sigma^>_L\mathbf G^{a}
    - \mathbf\Sigma^{>}_{\lambda} \mathbf G^{r}\mathbf\Sigma^<_L \mathbf G^{a}]( \varepsilon),\quad
\label{eq.Iinel}
\end{eqnarray}
where $\mathrm{G}_0=2e^2/h$ is the conductance quantum and summation over the vibration index $\lambda$ is assumed.
The e-vib self-energies ${\bf \Sigma}_{\lambda}$ are expressed as
\begin{eqnarray}
{\bf \Sigma}_{\lambda}^{\gtrless}( \varepsilon)= {\bf M}_{\lambda}\!\left\{
(N_\lambda+1){\bf G}^{\gtrless}(\varepsilon_\mp)+ N_\lambda{\bf
G}^{\gtrless}( \varepsilon_\pm) \right\}\!{\bf M}_{\lambda},
\quad\label{eq.sigmalesserph}\\
{\bf \Sigma}_{\lambda}^{r,a}(\varepsilon) = \pm \frac{1}{2}\left\{{\bf\Sigma}_\lambda^>(\varepsilon)
-{\bf\Sigma}_\lambda^<(\varepsilon)\right\}\label{eq.Resret}
-\frac{i}{2}\mathcal{H}\left[{\bf\Sigma}_\lambda^>
-{\bf\Sigma}_\lambda^<\right](\varepsilon)\label{eq.Imsret},
\end{eqnarray}
with $\varepsilon_\pm=\varepsilon\pm \hbar\omega_\lambda$, bosonic occupations $N_\lambda$, and $\mathcal H$ denoting the Hilbert transform.
Finally, the lesser/greater Green's functions ${\bf G}^\lessgtr$ describing the occupied/unoccupied states,
\begin{equation}
{\bf G}^{\gtrless}(\varepsilon)=\mp i\left\{ f_L(\mp\varepsilon){\bf A}_L(\varepsilon) + f_R(\mp\varepsilon){\bf A}_R(\varepsilon)\right\}\,,
\label{eq.Ggtrless}
\end{equation}
are given by the spectral density matrices ${\bf A}_\alpha=\mathbf G^r \mathbf \Gamma_\alpha \mathbf G^a$ for left/right moving 
states with fillings according to the reservoir Fermi-functions, $f_{\alpha}(\varepsilon)=n_F(\varepsilon-\mu_\alpha)$. 

The above equations are numerically demanding because of the energy integration over \emph{voltage-dependent} traces.
In the following we describe how further simplifications are possible without resorting to the WBA. 
We are here interested in the ``vibration-signal'', that is the change in the current close to the excitation threshold, $|eV|\approx\hbar\omega_\lambda$, with $eV=\mu_L-\mu_R$. 
As IETS signals are obtained at low temperatures, we assume that 
this is the smallest energy scale, $k_B T \ll \hbar\omega_\lambda,\Gamma$, where $\Gamma$ is 
the typical electronic resonance broadening.
The inelastic term $I_i$  [\Eqref{eq.Iinel}] then reduces to 
\begin{eqnarray}
	{I}_i &\approx& \frac{\mathrm{G}_0}{2e}\! \sum_{\sigma=\pm}\left(\coth\frac{\hbar\omega_\lambda}{2k_BT}-\coth\frac{\hbar\omega_\lambda+\sigma eV}{2k_BT}\right)\\
	&\times&\int_{-\infty}^\infty\!{d\varepsilon}{\rm Tr}\left[\mathbf M_\lambda \tilde{\mathbf A}_L(\varepsilon) \mathbf M_\lambda 
	\mathbf A_{R}(\varepsilon_\sigma) \right]\!\{f_L(\varepsilon)-f_{R}(\varepsilon_\sigma)\},\!
	\nonumber
\end{eqnarray}
where $\tilde{\mathbf A}_\alpha=\mathbf G^a \mathbf \Gamma_\alpha \mathbf G^r$ is the time-reversed version of $\mathbf A_\alpha$.
In the 2nd derivative of $I_i$ w.r.t. voltage $V$, the coth-parts give rise to a
sharply peaked signal around $|eV|=\hbar\omega_\lambda$ with width of the order of
$k_BT$. If the electronic structure ($\mathbf A_\alpha$) varies slowly on the
$k_BT$ scale, it can be replaced by a constant using $\varepsilon\approx \mu_L$
and $\varepsilon_\sigma\approx \mu_R=\mu_L+\sigma\hbar\omega_\lambda$. Thus, around the
vibration threshold we get  
\begin{eqnarray}
	\label{eq:ic4af}
	\partial_{V}^2{I}_i &\approx& \gamma_{i,\lambda}\, \partial_{V}^2 \mathcal{I}^\mathrm{sym},\\
	\gamma_{i,\lambda}&=&{\rm Tr}\!\left[ \mathbf M_\lambda \tilde{\mathbf A}_L(\mu_L) \mathbf M_\lambda \mathbf A_R(\mu_R) \right], \label{eq:gamma-i}
\end{eqnarray}
where we, as in the LOE-WBA, define the ``universal'' function
\begin{eqnarray}
	\mathcal{I}^\mathrm{sym}\! &\equiv&\! \frac{\mathrm{G}_0}{2e}\sum_{\sigma=\pm} \!\!\!\sigma(\hbar\omega_\lambda+\sigma eV)\\
&&\times\!\left(\!\coth\!\frac{\hbar\omega_ \lambda}{2k_BT}-\coth\!\frac{\hbar\omega_\lambda+\sigma eV}{2k_BT}\!\right)\nonumber.
\end{eqnarray}

The elastic term $I_e$ [\Eqref{eq.Iel}] can be divided in two parts, $I_e=I_e^n+I_e^h$, 
where the first(latter) represents all terms without(with) the Hilbert transformation originating in \Eqref{eq.Imsret}.
The ``non-Hilbert'' part $I_e^n$ yields a coth-factor and integral of similar in form to the one for $I_i$. Both $I_i$ and $I_e^n$ thus yield an inelastic signal with a lineshape 
given by the function $\partial_{V}^2 \mathcal{I}^\mathrm{sym}$
and the sign/intensity governed by $\gamma_\lambda=\gamma_{i,\lambda}+\gamma_{e,\lambda}$, with $\gamma_{e,\lambda}\approx{\rm Im} B_\lambda$, and
\begin{widetext}
\begin{eqnarray}
B_\lambda &\equiv&{\rm Tr}[\mathbf M_\lambda\mathbf A_R(\mu_L)\mathbf\Gamma_L(\mu_L)\mathbf G^r(\mu_L)\mathbf M_\lambda\mathbf A_R(\mu_R)
-\mathbf M_\lambda \mathbf G^a(\mu_R)\mathbf\Gamma_L(\mu_R)\mathbf A_R(\mu_R)\mathbf M_\lambda \mathbf A_L(\mu_L)]. \label{eq:ic4ag}
\end{eqnarray}
\end{widetext}
The ``Hilbert'' part $I_e^h$ requires a bit more consideration. Besides terms which do not result in threshold signals \cite{Haupt2010}, we have terms involving $\mathcal{H}[\mathbf A_\alpha f_\alpha]$. 
Again, if $\mathbf A_\alpha$ varies slowly around the step in $f_\alpha$ we may approximate
\begin{equation}
\mathcal{H}[\mathbf A_\alpha(\varepsilon') 
f_\alpha(\varepsilon')](\varepsilon) \approx \mathbf A_\alpha(\varepsilon) \mathcal{H}[f_\alpha(\varepsilon')](\varepsilon)\,. 
\end{equation}
The Hilbert transformation of the Fermi function is strongly peaked at the chemical potential, and again we evaluate the energy integral by evaluating all electronic structure functions
($\mathbf A_\alpha, \mathbf G^r,\mathbf \Gamma_\alpha$) at the peak values, keeping only the energy dependence of the functions related to $f_\alpha$ inside the integral.
The result is
\begin{equation}
\partial_{V}^2{I}_e^h\approx\kappa_{\lambda}\,\partial_{V}^2 \mathcal{I}^\mathrm{asym},
\end{equation}
with $\kappa_{\lambda}=2{\rm Re} B_\lambda$ and, again as in the LOE-WBA, the ``universal'' function
\begin{widetext}
\begin{eqnarray}
	\mathcal{I}^\mathrm{asym} &\equiv& \frac{\mathrm{G}_0}{2e}\int_{-\infty}^{+\infty}\!d\varepsilon\mathcal{H}\{ f(\varepsilon'_-)- f(\varepsilon'_+)\}(\varepsilon)\left(f(\varepsilon-eV)-f_{}(\varepsilon)\right)
	\approx-\frac{G_0}{2e\pi}\sum_{\sigma=\pm} \sigma|eV+\sigma\hbar\omega_\lambda|{\rm ln}\left| \frac{eV+\sigma\hbar\omega_\lambda}{\hbar\omega_\lambda} \right| \,.
	\label{eq:ass}
\end{eqnarray}
\end{widetext}
Here the latter is for $k_BT=0$, while it can be expressed using the digamma function for finite $k_BT$ \cite{Bevilacqua2013}. 
In total we have written the IETS as a sum of individual vibration signals \cite{PaFrBr.05.Modelinginelasticphonona},
\begin{eqnarray}
\label{eq:current}
\partial_{V}^2{I}(V)&=&  \gamma_\lambda \, \partial_{V}^2{\cal I}^\mathrm{sym}(V,\hbar\omega_\lambda, T, N_\lambda)\\
 &&+  \kappa_\lambda\,\partial_{V}^2{\cal I}^\mathrm{asym}(V,\hbar\omega_\lambda, T).\nonumber
\end{eqnarray}
Equation (\ref{eq:current}) is our main formal result.
As for the LOE-WBA we have expressed the vibration signals from the ``universal'' functions, and structure factors containing quantities 
readily obtained from DFT-NEGF. However, importantly, we have here generalized these to include
the effect of {\em finite} $\hbar\omega_\lambda$, and thus the change in electronic structure over the excitation energy.
Our LOE expressions for $\gamma_\lambda$ and $\kappa_\lambda$ above simply reduce to the LOE-WBA when $\mu_L=\mu_R=\mu_0$.
We will now demonstrate some situations where the LOE expression \Eqref{eq:current} is crucial for detailed interpretation of experimental IETS lineshapes.

\begin{figure}[!ht]
\includegraphics[scale=1.0]{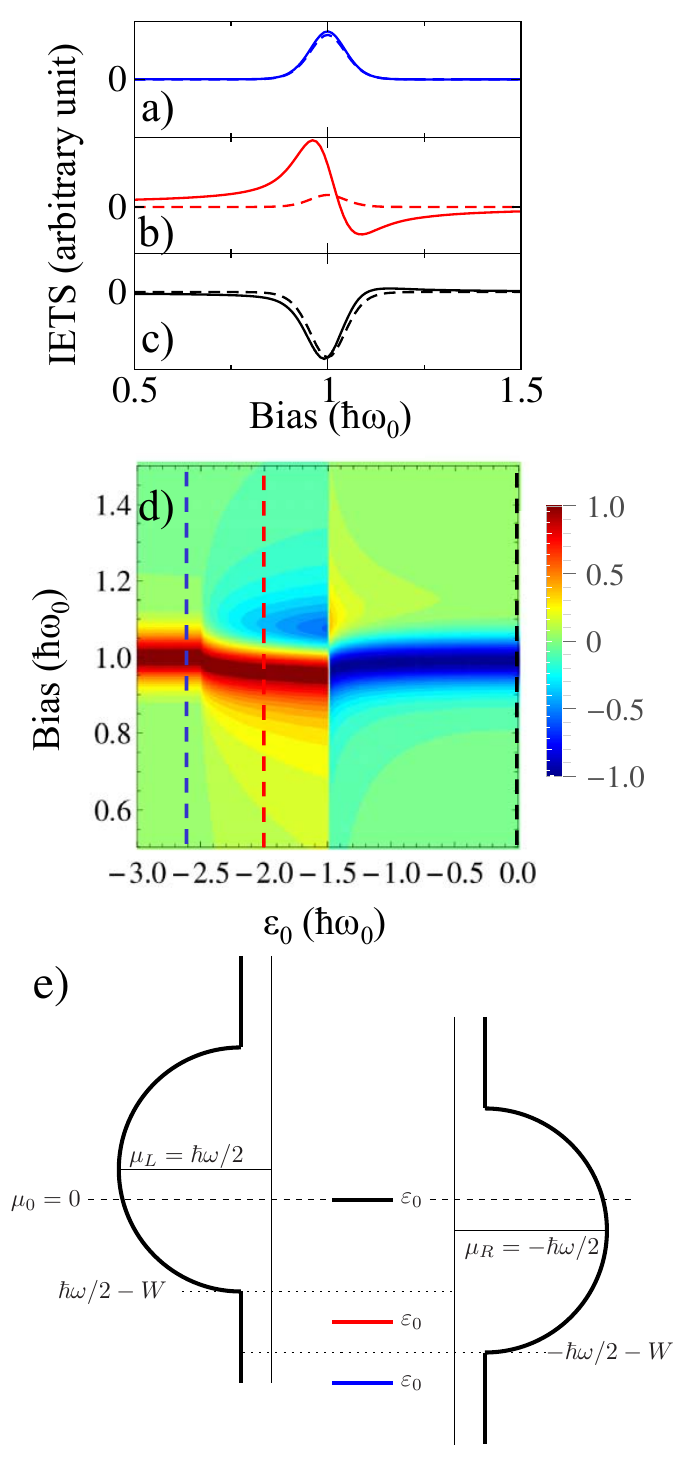}
\caption{(color online) (a-c) IETS spectrum from LOE (solid) and LOE-WBA (dashed) for three different position of an electronic
level, coupling with a wide $s$ band with constant density of states, and a
narrow $d$ band with bandwidth $W$ centered at the equilibrium Fermi level
$\mu_0=0$. (a) $\varepsilon_0 = -2.6$, (b) $\varepsilon_0=-2$, (c)
$\varepsilon_0=0$.  The transmission coefficients at the Fermi level are $T=0.006,
0.01, 1$, respectively.  (d) Contour plot of the IETS spectrum for different level positions. The signal is normalized such that
for each given $\varepsilon_0$, the height of the largest peak or dip is 1.
Parameters in unit of the vibration energy $\hbar\omega_0$: $t'=2t=W=2$,
$\Gamma_0=0.1$, $k_BT = 0.02$. Here, $t$ is the hopping matrix element of the $d$ band, and
$t'$ is its coupling to the electronic level.
(e) Schematics of the one-level model (shown for the three different level positions)
biased at the emission threshold $V=\hbar\omega$.} 
\label{fig:edepmu}
\end{figure}

{\it Simple model.} First we use a single-level model to illustrate how the ``asymmetric'' term contains important information 
about the energy dependence of the electrode couplings. In the LOE-WBA one
always has $\kappa_{\lambda}=0$ for symmetric junctions. This is \emph{not} the
case for the LOE expression \Eqref{eq:ic4ag}.  We therefore consider a {\em
symmetric junction} containing a single electronic level at $\varepsilon_0$
(with $\mu_0=0$), coupled to a local vibration ($\omega_0$), and with
energy-dependent electrode coupling rates.  Assuming symmetrical potential
drop, and using the notations $\Gamma_l=\Gamma_L(\mu_L)+\Gamma_R(\mu_L)$ and
$\Gamma_r=\Gamma_L(\mu_R)+\Gamma_R(\mu_R)$ we can write the ``symmetric'',
\begin{equation}
	\gamma =  -C\left\{ \Gamma_l^2\Gamma_r^2-
	\left(  4\varepsilon_0^2-\hbar^2\omega_0^2 \right)^2\right\},\label{eq:simple-model-sym}
\end{equation}
and ``asymmetric'' coefficients,
\begin{equation}
	\kappa =4C\left(\delta\Gamma\, \varepsilon_0+\bar\Gamma\,\hbar\omega_0\right)\left\{ \Gamma_l\Gamma_r-\left(4\varepsilon_0^2-\hbar^2\omega_0^2\right)\right\},
	\label{eq.kappamodel}
\end{equation}
where $\delta \Gamma=\Gamma_l-\Gamma_r$, $\bar\Gamma=(\Gamma_l+\Gamma_r)/2$,
and $C$ a constant common to $\gamma$ and $\kappa$. 
In the typical case of transition metal electrodes the coupling can contain contributions both from a wide $s$-band as well as from a 
narrow $d$-band leading to a significant $\delta\Gamma$ and finite $\kappa$.  To model the $s$-band we use a constant $\Gamma_0$, and 
to mimic the coupling (hopping $t'$) to a $d$-band we add the self-energy of a semi-infinite 1D chain, with bandwidth $2W$ centered at $\mu_0=0$.
Figure~\ref{fig:edepmu} (a-c) compares the signals calculated from LOE-WBA and LOE for different $\varepsilon_0$. 
For both treatments we observe that the peak in the off-resonance IETS evolves into a dip on-resonance. However, only in the LOE the two regimes 
are separated by a peak-dip structure close to resonance due to the ``asymmetric'' $\kappa$, which is enhanced at the 
onset of the coupling with $d$ band in one electrode. The change in IETS signal with a gate-potential ($\varepsilon_0$) is shown in Fig.~\ref{fig:edepmu} (d).
The features observed at $\varepsilon_0=\pm\hbar\omega/2-W$ is associated with the level being resonant with the left/right $d$-band onset, respectively.

\begin{figure}[thpb]
\begin{center}
\includegraphics[width=0.45\textwidth]{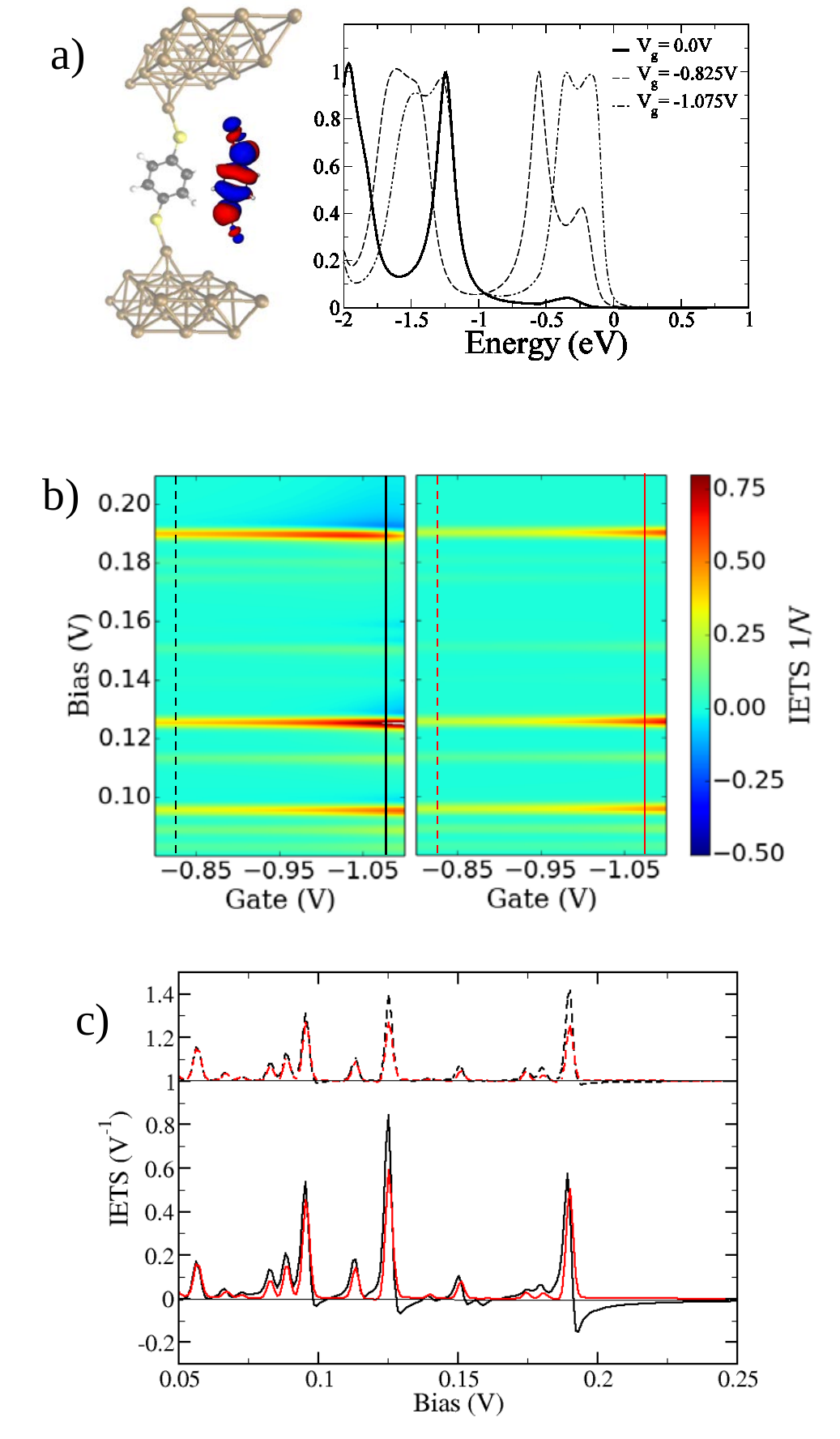}
\end{center}
\caption{(color online) (a) BDT between two adatoms on Au(111) together with transmission 
for off resonance (zero gate) and close to resonance.  (b) IETS as a function of gate voltage from
LOE (left) and LOE-WBA (right). (c) IETS for fixed gate voltage off-resonance (dashed lines, offset for clarity) 
and close-to-resonance (solid lines). Black: LOE, Red: LOE-WBA.
The IETS signals are calculated for $T = 4.2$ K and processed to mimic the 
experimental broadening arising from the lock-in technique with a harmonic voltage 
modulation of $V_\mathrm{rms} = 5$ mV \cite{Frederiksen2007}.
}
\label{fig:bdt}
\end{figure}

{\it IETS of benzene-dithiol.}
It has been possible to apply an external gate potential to junctions with
small molecules between metallic
electrodes \cite{YuKeCi.04.Inelasticelectrontunneling,Song2009}. Under these conditions
IETS have been recorded for gated octane-dithiol (ODT) and benzene-dithiol (BDT)
molecules between gold electrodes \cite{Song2009}.
For both ODT and BDT the
quite symmetric \IV~characteristics indicates a symmetric bonding to the
electrodes. For the $\pi$-conjugated BDT it was shown how the transport can be
tuned from far off-resonance ($G\sim 0.01 G_0$) to close to the HOMO resonance
increasing the conductance by more than an order of magnitude.  As in the
simple symmetric model above, this was reflected in the shape of the IETS
signal for BDT going from a peak for off-resonance, to a peak-dip close to
resonance, with the peaks appearing at the same voltages.  However,
the analysis by Song \etal~\cite{Song2009} was based on a model assuming asymmetric electrode
couplings at zero bias (STM regime) \cite{PEBA.87.INELASTICELECTRON-TUNNELINGFROM}. 
Our simple model [Fig.~1(b)] instead suggests that the observed peak-dip lineshape originates solely 
from the $\Gamma_{l},\Gamma_{r}$ asymmetry driven by the bias voltage near resonance
rather than from asymmetric electrode couplings ($\Gamma_{L},\Gamma_{R}$).

Next, we turn to our DFT-NEGF calculations \footnote{We employ the
SIESTA \cite{SoArGa.02.SIESTAmethodab}/TranSIESTA \cite{BrMoOr.02} method with
the GGA-PBE \cite{PeBuEr.96} exchange-correlation functional.
Electron-vibration couplings and IETS are calculated with
Inelastica \cite{Frederiksen2007}.}.  The importance of an efficient
scheme is underlined by the fact that an IETS calculation is required for each
gate value. In the break-junction experiments the atomic structure of the
junction is unknown. We anticipate that the gap between the electrodes is quite
open and involves sharp asperities with low-coordinated gold atoms in order to
allow for the external gating to be effective. In order to emulate this, we
consider BDT bonded between adatoms on Au(111) surfaces
[Fig.~\ref{fig:bdt}(a)], and employ only the $\Gamma$-point in the transport
calculations yielding sharper features in the electronic structure. We correct
the HOMO-LUMO gap \cite{Garcia-Suarez2011} and model the electrostatic gating
simply by a rigid shift of the molecular orbital energies relative to the gold
energies. In Fig.~\ref{fig:bdt}(b)-(c) we compare IETS calculated with LOE and
LOE-WBA as a function of gating. 
As in the experiment, we observe three clear signals around
$\hbar\omega=95, 130, 200$ meV due to benzene vibrational modes. 
Off resonance, the LOE and LOE-WBA are in agreement as expected. But when the
gate voltage is tuned to around $V_g\approx -1$ V the methods deviate because
of the appearance of sharp resonances in the transmission around the Fermi
energy [Fig.~\ref{fig:bdt}(a)]. These resonances involve the $d$-orbitals on
the contacting gold atoms, as seen in the eigenchannel \cite{Paulsson07} plot in
Fig.~\ref{fig:bdt}(a), and result in a peak-dip structure as
seen in the experiment and anticipated by the simple model. 
Thus it is important to go beyond LOE-WBA in order to reproduce 
the peak to peak-dip transition taken as evidence for close-to-resonance transport.

\begin{figure}[thpb]
  \includegraphics[width=\columnwidth]{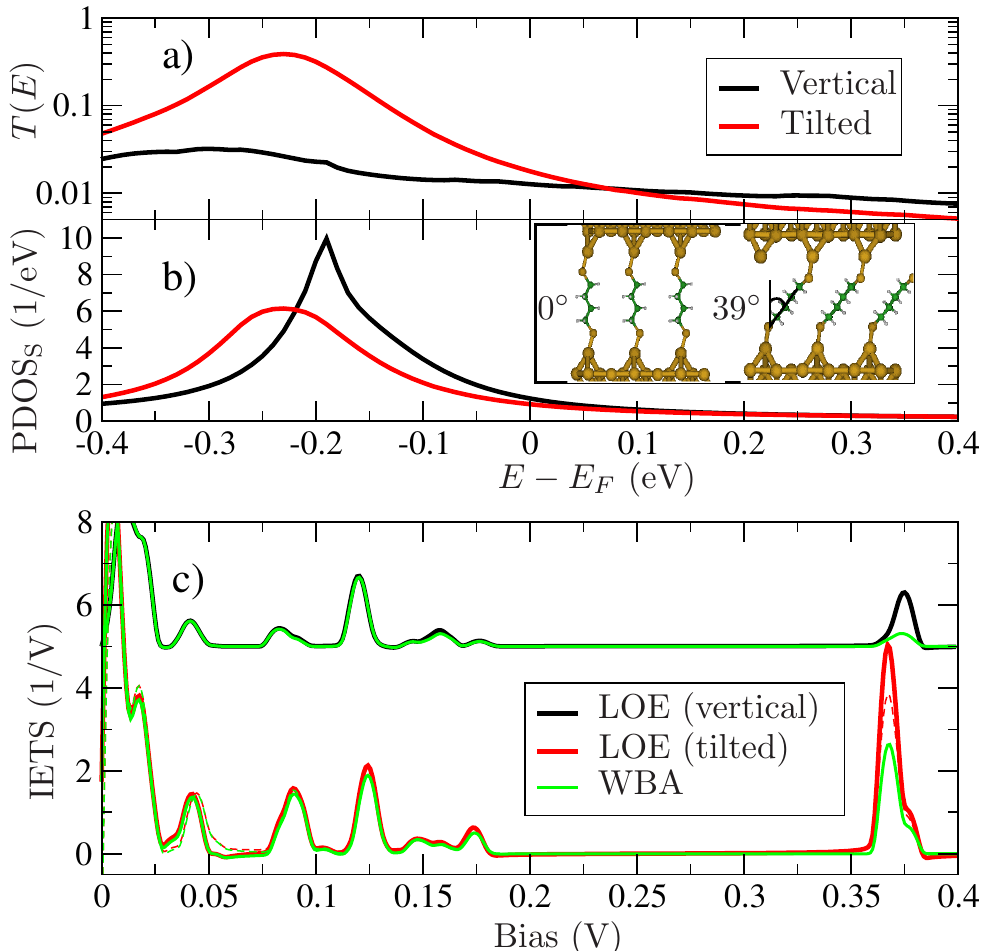}
  \caption{(color online) (a) Transmission and (b) projected density of states over 
S for vertical and tilted C4DT in a $2\times 2$ supercell of Au(111). (c) IETS 
within LOE and LOE-WBA (averaged over $k_{||}$) using $T=4.2$ K and $V_\mathrm{rms}=5$ mV \cite{Frederiksen2007}.
Thin dashed lines represent the reverse bias polarity.}
\label{fig:T-PDOS-IETS}
\end{figure}

{\it IETS of alkane-dithiol.}
As another demonstration of the improvement of LOE over LOE-WBA, we consider
molecular junctions formed by straight or tilted butane-dithiol (C4DT) molecules linked via low-coordinated
Au adatoms to Au(111) electrodes, see inset to Fig.~\ref{fig:T-PDOS-IETS}.
Based on DFT-NEGF [27] we calculate elastic transmission and IETS for the periodic
structure averaged over electron momentum $k_{||}$ \cite{Foti2014}.
As shown in Fig.~\ref{fig:T-PDOS-IETS}(a)-(b), transport around the Fermi level 
is off-resonance but dominated by the tail of a sulfur-derived peak centered at approximately 
0.25 eV below the Fermi level. This feature introduces a relatively strong energy-dependence 
into the electronic structure which makes the WBA questionable.
Indeed, as shown in Fig.~\ref{fig:T-PDOS-IETS}(c), LOE-WBA gives a smaller
IETS intensity compared to the LOE for the energetic CH$_{2}$ stretch modes ($\hbar\omega\sim 375$ meV). 
The WBA may thus be the reason why LOE-WBA calculations was reported to underestimate the IETS intensity 
for these energetic modes in comparison with experiments \cite{Okabayashi2010}.
We note that the intensity enhancement is found to be more pronounced for the 
straight configuration,
which may be rationalized from \Eqref{eq:simple-model-sym} (the condition 
$\varepsilon_0=\pm\hbar\omega_0/2$ better describes the vertical than the tilted case).
The intensity change reported in Fig.~\ref{fig:T-PDOS-IETS} 
thus suggests the relevance of going beyond LOE-WBA for simulations involving
high-energy vibrational modes.

{\it Conclusions.}
A generalized LOE scheme for IETS simulations with the DFT-NEGF method has been described. Without
introducing the WBA, our formulation retains both the transparency and computational efficiency of the LOE-WBA.
This improvement is important to capture correctly the IETS lineshape in situations where the electronic
structure varies appreciably on the scale of the vibration energies, such as near
sharp resonances or band edges. Together with DFT-NEGF calculations 
we have discovered that the intricate experimental lineshape of a gated BDT can be
explained without the need to assume asymmetric bonding of the molecule to the electrodes.
Also, simulations for C4DT junctions suggest that going beyond WBA is
important to capture the IETS intensity related to energetic CH$_{2}$ 
stretch modes.

We acknowledge computer resources from the DCSC.  JTL acknowledges support from
the National Natural Science Foundation of China (Grant No.~11304107, 61371015), 
and the Fundamental Research Funds for the Central Universities (HUST:2013TS032).  
GF and TF acknowledge support from the Basque Departamento de
Educac\'ion and the UPV/EHU (Grant No.~IT-756-13), the Spanish Ministerio de
Econom\'ia y Competitividad (Grant No.~FIS2010-19609-CO2-00), and the European
Union Integrated Project PAMS.

\bibliography{ExLOEBibliography}

\begin{thebibliography}{26}
\expandafter\ifx\csname natexlab\endcsname\relax\def\natexlab#1{#1}\fi
\expandafter\ifx\csname bibnamefont\endcsname\relax
  \def\bibnamefont#1{#1}\fi
\expandafter\ifx\csname bibfnamefont\endcsname\relax
  \def\bibfnamefont#1{#1}\fi
\expandafter\ifx\csname citenamefont\endcsname\relax
  \def\citenamefont#1{#1}\fi
\expandafter\ifx\csname url\endcsname\relax
  \def\url#1{\texttt{#1}}\fi
\expandafter\ifx\csname urlprefix\endcsname\relax\def\urlprefix{URL }\fi
\providecommand{\bibinfo}[2]{#2}
\providecommand{\eprint}[2][]{\url{#2}}

\bibitem[{\citenamefont{Stipe et~al.}(1998)\citenamefont{Stipe, Rezaei, and
  Ho}}]{StReHo.98.Single-moleculevibrationalspectroscopy}
\bibinfo{author}{\bibfnamefont{B.~C.} \bibnamefont{Stipe}},
  \bibinfo{author}{\bibfnamefont{M.~A.} \bibnamefont{Rezaei}},
  \bibnamefont{and} \bibinfo{author}{\bibfnamefont{W.}~\bibnamefont{Ho}},
  \bibinfo{journal}{Science} \textbf{\bibinfo{volume}{280}},
  \bibinfo{pages}{1732} (\bibinfo{year}{1998}).

\bibitem[{\citenamefont{Agrait et~al.}(2002)\citenamefont{Agrait, Untiedt,
  Rubio-Bollinger, and Vieira}}]{AgUnRu.02.Onsetofenergy}
\bibinfo{author}{\bibfnamefont{N.}~\bibnamefont{Agrait}},
  \bibinfo{author}{\bibfnamefont{C.}~\bibnamefont{Untiedt}},
  \bibinfo{author}{\bibfnamefont{G.}~\bibnamefont{Rubio-Bollinger}},
  \bibnamefont{and} \bibinfo{author}{\bibfnamefont{S.}~\bibnamefont{Vieira}},
  \bibinfo{journal}{Phys.~Rev.~Lett.} \textbf{\bibinfo{volume}{88}},
  \bibinfo{pages}{216803} (\bibinfo{year}{2002}).

\bibitem[{\citenamefont{Smit et~al.}(2002)\citenamefont{Smit, Noat, Untiedt,
  Lang, van Hemert, and van Ruitenbeek}}]{SmNoUn.02.Measurementofconductance}
\bibinfo{author}{\bibfnamefont{R.~H.~M.} \bibnamefont{Smit}},
  \bibinfo{author}{\bibfnamefont{Y.}~\bibnamefont{Noat}},
  \bibinfo{author}{\bibfnamefont{C.}~\bibnamefont{Untiedt}},
  \bibinfo{author}{\bibfnamefont{N.~D.} \bibnamefont{Lang}},
  \bibinfo{author}{\bibfnamefont{M.~C.} \bibnamefont{van Hemert}},
  \bibnamefont{and} \bibinfo{author}{\bibfnamefont{J.~M.} \bibnamefont{van
  Ruitenbeek}}, \bibinfo{journal}{Nature (London)}
  \textbf{\bibinfo{volume}{419}}, \bibinfo{pages}{906} (\bibinfo{year}{2002}).

\bibitem[{\citenamefont{Kushmerick et~al.}(2004)\citenamefont{Kushmerick,
  Lazorcik, Patterson, Shashidhar, Seferos, and
  Bazan}}]{KuLaPa.04.Vibroniccontributionsto}
\bibinfo{author}{\bibfnamefont{J.~G.} \bibnamefont{Kushmerick}},
  \bibinfo{author}{\bibfnamefont{J.}~\bibnamefont{Lazorcik}},
  \bibinfo{author}{\bibfnamefont{C.~H.} \bibnamefont{Patterson}},
  \bibinfo{author}{\bibfnamefont{R.}~\bibnamefont{Shashidhar}},
  \bibinfo{author}{\bibfnamefont{D.~S.} \bibnamefont{Seferos}},
  \bibnamefont{and} \bibinfo{author}{\bibfnamefont{G.~C.} \bibnamefont{Bazan}},
  \bibinfo{journal}{Nano Lett.} \textbf{\bibinfo{volume}{4}},
  \bibinfo{pages}{639} (\bibinfo{year}{2004}).

\bibitem[{\citenamefont{Yu et~al.}(2004)\citenamefont{Yu, Keane, Ciszek, Cheng,
  Stewart, Tour, and Natelson}}]{YuKeCi.04.Inelasticelectrontunneling}
\bibinfo{author}{\bibfnamefont{L.~H.} \bibnamefont{Yu}},
  \bibinfo{author}{\bibfnamefont{Z.~K.} \bibnamefont{Keane}},
  \bibinfo{author}{\bibfnamefont{J.~W.} \bibnamefont{Ciszek}},
  \bibinfo{author}{\bibfnamefont{L.}~\bibnamefont{Cheng}},
  \bibinfo{author}{\bibfnamefont{M.~P.} \bibnamefont{Stewart}},
  \bibinfo{author}{\bibfnamefont{J.~M.} \bibnamefont{Tour}}, \bibnamefont{and}
  \bibinfo{author}{\bibfnamefont{D.}~\bibnamefont{Natelson}},
  \bibinfo{journal}{Phys.~Rev.~Lett.} \textbf{\bibinfo{volume}{93}},
  \bibinfo{pages}{266802} (\bibinfo{year}{2004}).

\bibitem[{\citenamefont{Song et~al.}(2009)\citenamefont{Song, Kim, Jang, Jeong,
  Reed, and Lee}}]{Song2009}
\bibinfo{author}{\bibfnamefont{H.}~\bibnamefont{Song}},
  \bibinfo{author}{\bibfnamefont{Y.}~\bibnamefont{Kim}},
  \bibinfo{author}{\bibfnamefont{Y.~H.} \bibnamefont{Jang}},
  \bibinfo{author}{\bibfnamefont{H.}~\bibnamefont{Jeong}},
  \bibinfo{author}{\bibfnamefont{M.~A.} \bibnamefont{Reed}}, \bibnamefont{and}
  \bibinfo{author}{\bibfnamefont{T.}~\bibnamefont{Lee}},
  \bibinfo{journal}{Nature} \textbf{\bibinfo{volume}{462}},
  \bibinfo{pages}{1039} (\bibinfo{year}{2009}).

\bibitem[{\citenamefont{Okabayashi et~al.}(2013)\citenamefont{Okabayashi,
  Paulsson, and Komeda}}]{Okabayashi2013}
\bibinfo{author}{\bibfnamefont{N.}~\bibnamefont{Okabayashi}},
  \bibinfo{author}{\bibfnamefont{M.}~\bibnamefont{Paulsson}}, \bibnamefont{and}
  \bibinfo{author}{\bibfnamefont{T.}~\bibnamefont{Komeda}},
  \bibinfo{journal}{Prog. Surf. Sci.} \textbf{\bibinfo{volume}{88}},
  \bibinfo{pages}{1} (\bibinfo{year}{2013}).

\bibitem[{\citenamefont{Galperin et~al.}(2007)\citenamefont{Galperin, Ratner,
  and Nitzan}}]{GaRaNi07}
\bibinfo{author}{\bibfnamefont{M.}~\bibnamefont{Galperin}},
  \bibinfo{author}{\bibfnamefont{M.~A.} \bibnamefont{Ratner}},
  \bibnamefont{and} \bibinfo{author}{\bibfnamefont{A.}~\bibnamefont{Nitzan}},
  \bibinfo{journal}{J.~Phys.: Condens.~Matter} \textbf{\bibinfo{volume}{19}},
  \bibinfo{pages}{103201} (\bibinfo{year}{2007}).

\bibitem[{\citenamefont{Sergueev et~al.}(2005)\citenamefont{Sergueev, Roubtsov,
  and Guo}}]{SeRoGu.05.Abinitioanalysis}
\bibinfo{author}{\bibfnamefont{N.}~\bibnamefont{Sergueev}},
  \bibinfo{author}{\bibfnamefont{D.}~\bibnamefont{Roubtsov}}, \bibnamefont{and}
  \bibinfo{author}{\bibfnamefont{H.}~\bibnamefont{Guo}},
  \bibinfo{journal}{Phys.~Rev.~Lett.} \textbf{\bibinfo{volume}{95}},
  \bibinfo{pages}{146803} (\bibinfo{year}{2005}).

\bibitem[{\citenamefont{Paulsson et~al.}(2005)\citenamefont{Paulsson,
  Frederiksen, and Brandbyge}}]{PaFrBr.05.Modelinginelasticphonona}
\bibinfo{author}{\bibfnamefont{M.}~\bibnamefont{Paulsson}},
  \bibinfo{author}{\bibfnamefont{T.}~\bibnamefont{Frederiksen}},
  \bibnamefont{and}
  \bibinfo{author}{\bibfnamefont{M.}~\bibnamefont{Brandbyge}},
  \bibinfo{journal}{Phys.~Rev.~B} \textbf{\bibinfo{volume}{72}},
  \bibinfo{pages}{201101} (\bibinfo{year}{2005}).

\bibitem[{\citenamefont{Jiang et~al.}(2005)\citenamefont{Jiang, Kula, Lu, and
  Luo}}]{JiKuLu.05.First-principlessimulationsof}
\bibinfo{author}{\bibfnamefont{J.}~\bibnamefont{Jiang}},
  \bibinfo{author}{\bibfnamefont{M.}~\bibnamefont{Kula}},
  \bibinfo{author}{\bibfnamefont{W.}~\bibnamefont{Lu}}, \bibnamefont{and}
  \bibinfo{author}{\bibfnamefont{Y.}~\bibnamefont{Luo}}, \bibinfo{journal}{Nano
  Lett.} \textbf{\bibinfo{volume}{5}}, \bibinfo{pages}{1551}
  (\bibinfo{year}{2005}).

\bibitem[{\citenamefont{Solomon et~al.}(2006)\citenamefont{Solomon, Gagliardi,
  Pecchia, Frauenheim, Di~Carlo, Reimers, and
  Hush}}]{SoGaPe.06.Understandinginelasticelectron-tunneling}
\bibinfo{author}{\bibfnamefont{G.~C.} \bibnamefont{Solomon}},
  \bibinfo{author}{\bibfnamefont{A.}~\bibnamefont{Gagliardi}},
  \bibinfo{author}{\bibfnamefont{A.}~\bibnamefont{Pecchia}},
  \bibinfo{author}{\bibfnamefont{T.}~\bibnamefont{Frauenheim}},
  \bibinfo{author}{\bibfnamefont{A.}~\bibnamefont{Di~Carlo}},
  \bibinfo{author}{\bibfnamefont{J.~R.} \bibnamefont{Reimers}},
  \bibnamefont{and} \bibinfo{author}{\bibfnamefont{H.~S.} \bibnamefont{Hush}},
  \bibinfo{journal}{J.~Chem.~Phys.} \textbf{\bibinfo{volume}{124}},
  \bibinfo{pages}{094704} (\bibinfo{year}{2006}).

\bibitem[{\citenamefont{Frederiksen et~al.}(2007)\citenamefont{Frederiksen,
  Paulsson, Brandbyge, and Jauho}}]{Frederiksen2007}
\bibinfo{author}{\bibfnamefont{T.}~\bibnamefont{Frederiksen}},
  \bibinfo{author}{\bibfnamefont{M.}~\bibnamefont{Paulsson}},
  \bibinfo{author}{\bibfnamefont{M.}~\bibnamefont{Brandbyge}},
  \bibnamefont{and} \bibinfo{author}{\bibfnamefont{A.-P.} \bibnamefont{Jauho}},
  \bibinfo{journal}{Phys.~Rev.~B} \textbf{\bibinfo{volume}{75}},
  \bibinfo{pages}{205413} (\bibinfo{year}{2007}).

\bibitem[{\citenamefont{Rossen et~al.}(2013)\citenamefont{Rossen, Flipse, and
  Cerd\'a}}]{Rossen13}
\bibinfo{author}{\bibfnamefont{E.~T.~R.} \bibnamefont{Rossen}},
  \bibinfo{author}{\bibfnamefont{C.~F.~J.} \bibnamefont{Flipse}},
  \bibnamefont{and} \bibinfo{author}{\bibfnamefont{J.~I.}
  \bibnamefont{Cerd\'a}}, \bibinfo{journal}{Phys. Rev. B}
  \textbf{\bibinfo{volume}{87}}, \bibinfo{pages}{235412}
  (\bibinfo{year}{2013}).

\bibitem[{\citenamefont{Viljas et~al.}(2005)\citenamefont{Viljas, Cuevas,
  Pauly, and Hafner}}]{ViCuPa.05.Electron-vibrationinteractionin}
\bibinfo{author}{\bibfnamefont{J.~K.} \bibnamefont{Viljas}},
  \bibinfo{author}{\bibfnamefont{J.~C.} \bibnamefont{Cuevas}},
  \bibinfo{author}{\bibfnamefont{F.}~\bibnamefont{Pauly}}, \bibnamefont{and}
  \bibinfo{author}{\bibfnamefont{M.}~\bibnamefont{Hafner}},
  \bibinfo{journal}{Phys.~Rev.~B} \textbf{\bibinfo{volume}{72}},
  \bibinfo{pages}{245415} (\bibinfo{year}{2005}).

\bibitem[{\citenamefont{Persson and
  Baratoff}(1987)}]{PEBA.87.INELASTICELECTRON-TUNNELINGFROM}
\bibinfo{author}{\bibfnamefont{B.~N.~J.} \bibnamefont{Persson}}
  \bibnamefont{and} \bibinfo{author}{\bibfnamefont{A.}~\bibnamefont{Baratoff}},
  \bibinfo{journal}{Phys.~Rev.~Lett.} \textbf{\bibinfo{volume}{59}},
  \bibinfo{pages}{339} (\bibinfo{year}{1987}).

\bibitem[{\citenamefont{Egger and Gogolin}(2008)}]{Egger08}
\bibinfo{author}{\bibfnamefont{R.}~\bibnamefont{Egger}} \bibnamefont{and}
  \bibinfo{author}{\bibfnamefont{A.~O.} \bibnamefont{Gogolin}},
  \bibinfo{journal}{Phys. Rev. B} \textbf{\bibinfo{volume}{77}},
  \bibinfo{pages}{113405} (\bibinfo{year}{2008}).

\bibitem[{\citenamefont{Okabayashi et~al.}(2010)\citenamefont{Okabayashi,
  Paulsson, Ueba, Konda, and Komeda}}]{Okabayashi2010}
\bibinfo{author}{\bibfnamefont{N.}~\bibnamefont{Okabayashi}},
  \bibinfo{author}{\bibfnamefont{M.}~\bibnamefont{Paulsson}},
  \bibinfo{author}{\bibfnamefont{H.}~\bibnamefont{Ueba}},
  \bibinfo{author}{\bibfnamefont{Y.}~\bibnamefont{Konda}}, \bibnamefont{and}
  \bibinfo{author}{\bibfnamefont{T.}~\bibnamefont{Komeda}},
  \bibinfo{journal}{Nano Lett.} \textbf{\bibinfo{volume}{10}},
  \bibinfo{pages}{2950} (\bibinfo{year}{2010}).

\bibitem[{\citenamefont{Haupt et~al.}(2010)\citenamefont{Haupt, Novotny, and
  Belzig}}]{Haupt2010}
\bibinfo{author}{\bibfnamefont{F.}~\bibnamefont{Haupt}},
  \bibinfo{author}{\bibfnamefont{T.}~\bibnamefont{Novotny}}, \bibnamefont{and}
  \bibinfo{author}{\bibfnamefont{W.}~\bibnamefont{Belzig}},
  \bibinfo{journal}{Phys.~Rev.~B} \textbf{\bibinfo{volume}{82}},
  \bibinfo{pages}{165441} (\bibinfo{year}{2010}).

\bibitem[{\citenamefont{Bevilacqua}(2013)}]{Bevilacqua2013}
\bibinfo{author}{\bibfnamefont{G.}~\bibnamefont{Bevilacqua}}
  (\bibinfo{year}{2013}), \bibinfo{note}{arXiv:1303.6206 [math-ph]}.

\bibitem[{\citenamefont{García-Suárez and Lambert}(2011)}]{Garcia-Suarez2011}
\bibinfo{author}{\bibfnamefont{V.~M.} \bibnamefont{García-Suárez}}
  \bibnamefont{and} \bibinfo{author}{\bibfnamefont{C.~J.}
  \bibnamefont{Lambert}}, \bibinfo{journal}{New Journal of Physics}
  \textbf{\bibinfo{volume}{13}}, \bibinfo{pages}{053026}
  (\bibinfo{year}{2011}).

\bibitem[{\citenamefont{Paulsson and Brandbyge}(2007)}]{Paulsson07}
\bibinfo{author}{\bibfnamefont{M.}~\bibnamefont{Paulsson}} \bibnamefont{and}
  \bibinfo{author}{\bibfnamefont{M.}~\bibnamefont{Brandbyge}},
  \bibinfo{journal}{Phys. Rev. B} \textbf{\bibinfo{volume}{76}},
  \bibinfo{pages}{115117} (\bibinfo{year}{2007}).

\bibitem[{\citenamefont{Foti et~al.}()\citenamefont{Foti, Sanchez-Portal,
  Arnau, and Frederiksen}}]{Foti2014}
\bibinfo{author}{\bibfnamefont{G.}~\bibnamefont{Foti}},
  \bibinfo{author}{\bibfnamefont{D.}~\bibnamefont{Sanchez-Portal}},
  \bibinfo{author}{\bibfnamefont{A.}~\bibnamefont{Arnau}}, \bibnamefont{and}
  \bibinfo{author}{\bibfnamefont{T.}~\bibnamefont{Frederiksen}},
  \bibinfo{note}{(in preparation)}.

\bibitem[{\citenamefont{Soler et~al.}(2002)\citenamefont{Soler, Artacho, Gale,
  Garcia, Junquera, Ordejon, and Sanchez-Portal}}]{SoArGa.02.SIESTAmethodab}
\bibinfo{author}{\bibfnamefont{J.~M.} \bibnamefont{Soler}},
  \bibinfo{author}{\bibfnamefont{E.}~\bibnamefont{Artacho}},
  \bibinfo{author}{\bibfnamefont{J.~D.} \bibnamefont{Gale}},
  \bibinfo{author}{\bibfnamefont{A.}~\bibnamefont{Garcia}},
  \bibinfo{author}{\bibfnamefont{J.}~\bibnamefont{Junquera}},
  \bibinfo{author}{\bibfnamefont{P.}~\bibnamefont{Ordejon}}, \bibnamefont{and}
  \bibinfo{author}{\bibfnamefont{D.}~\bibnamefont{Sanchez-Portal}},
  \bibinfo{journal}{J.~Phys.: Condens.~Matter} \textbf{\bibinfo{volume}{14}},
  \bibinfo{pages}{2745} (\bibinfo{year}{2002}).

\bibitem[{\citenamefont{Brandbyge et~al.}(2002)\citenamefont{Brandbyge, Mozos,
  Ordejon, Taylor, and Stokbro}}]{BrMoOr.02}
\bibinfo{author}{\bibfnamefont{M.}~\bibnamefont{Brandbyge}},
  \bibinfo{author}{\bibfnamefont{J.~L.} \bibnamefont{Mozos}},
  \bibinfo{author}{\bibfnamefont{P.}~\bibnamefont{Ordejon}},
  \bibinfo{author}{\bibfnamefont{J.}~\bibnamefont{Taylor}}, \bibnamefont{and}
  \bibinfo{author}{\bibfnamefont{K.}~\bibnamefont{Stokbro}},
  \bibinfo{journal}{Phys.~Rev.~B} \textbf{\bibinfo{volume}{65}},
  \bibinfo{pages}{165401} (\bibinfo{year}{2002}).

\bibitem[{\citenamefont{Perdew et~al.}(1996)\citenamefont{Perdew, Burke, and
  Ernzerhof}}]{PeBuEr.96}
\bibinfo{author}{\bibfnamefont{J.~P.} \bibnamefont{Perdew}},
  \bibinfo{author}{\bibfnamefont{K.}~\bibnamefont{Burke}}, \bibnamefont{and}
  \bibinfo{author}{\bibfnamefont{M.}~\bibnamefont{Ernzerhof}},
  \bibinfo{journal}{Phys.~Rev.~Lett.} \textbf{\bibinfo{volume}{77}},
  \bibinfo{pages}{3865} (\bibinfo{year}{1996}).

\end{thebibliography}

\end{document}